# Superconducting properties of MgB$_2$ thin films prepared on flexible plastic substrates

Peter Kúš

*Department of Solid State Physics FMFI, Comenius University, SK-84248 Bratislava, Slovak Republic*

Andrej Plecenik and Leonid Satrapinsky

*Institute of Electrical Engineering, Slovak Academy of Science, SK-84239 Bratislava, Slovak Republic*

Ying Xu and Roman Sobolewski[a]

*Department of Electrical and Computer Engineering and the Laboratory for Laser Energetics, University of Rochester, Rochester, New York 14627-0231*

Superconducting MgB$_2$ thin films were prepared on 50-$\mu$m-thick, flexible polyamide Kapton-E foils by vacuum co-deposition of Mg and B precursors with nominal thickness of about 100 nm and a special *ex-situ* rapid annealing process in an Ar or vacuum atmosphere. In the optimal annealing process, the Mg-B films were heated to approximately 600°C, but at the same time, the backside of the structures was attached to a water-cooled radiator to avoid overheating of the plastic substrate. The resulting MgB$_2$ films were amorphous with the onset of the superconducting transition at $T_{c,\text{on}} \approx 33$ K and the transition width of approximately 3 K. The critical current density was $>7 \times 10^5$ A/cm$^2$ at 4.2 K, and its temperature dependence indicated a granular film composition with a network of intergranular weak links. The films could be deposited on large-area foils (up to 400 cm$^2$) and, after processing, cut into any shapes (e.g., stripes) with scissors or bent multiple times, without any observed degradation of their superconducting properties.

---

[a]Corresponding author: sobolewski@ece.rochester.edu. Also at the Institute of Physics, Polish Academy of Sciences, PL-02668 Warszawa, Poland.

The recent discovery of superconductivity at 39 K in hexagonal magnesium diborides[1] has stimulated very intensive investigations on the fundamental mechanism of superconductivity in $MgB_2$, as well as on possibilities of practical applications of this new superconductor. In comparison with high-temperature cuprates, $MgB_2$ superconductors have more than two times lower anisotropy, significantly larger coherence length, and higher transparency of grain boundaries to the current flow. At the same time, when compared to conventional metallic superconductors (including $Nb_3Sn$ or NbN), they have at least two times higher critical temperature $T_c$ and energy gap, as well as higher critical current density $j_c$. As a result, $MgB_2$ superconductors are expected to play an important role for high-current, high-field applications, as well as in cryoelectronics, where they might be the material-of-choice for above-300-GHz clock-rate digital electronics. In addition, $MgB_2$ devices could operate in simple cryocoolers.

The availability of high-quality superconducting thin films with single-crystal-like transport properties[2] is the key to realization of practical $MgB_2$ cryoelectronic devices. The films with high values of $T_c$ and $j_c$ have been prepared on $SrTiO_3$, $Al_2O_3$, Si, SiC, and other substrates by various deposition techniques.[3-8] All of the above procedures required post-annealing at temperatures higher than 600°C, and the superconducting properties of the resulting films depended very strongly on the conditions of their annealing. The best reported $MgB_2$ thin films are characterized by $T_c \approx 39$ K and $j_c > 10^7$ A/cm$^2$ at 4.2 K and $5 \times 10^5$ A/cm$^2$ at 30 K, both at zero external magnetic field.[9] Preparation of $MgB_2$ thin films on unconventional substrates, such as stainless steel[10] or plastic foils, is also desirable. $MgB_2$-on-steel films are interesting for preparation of low-magnetic shielding and antennas, while plastic foils are very attractive for their bending and shaping abilities.

In this letter, we present the preparation and superconducting properties of $MgB_2$ thin films fabricated on flexible substrates (Kapton-E polyamide foil), using three specially designed, rapid annealing processes. Flexible plastic substrates introduce a number of novel aspects to superconducting technology, such as the ability to cut the final structures into desired shapes for, e.g., magnetic shielding. They are also unbreakable and can be rolled into small superconducting coils or form flexible microwave or high-speed digital microstrips and coplanar transmission lines.

Kapton-E foils up to $20 \times 20$ cm$^2$ (limited by our deposition apparatus) were cleaned in acetone, ultrasonically washed in ethanol, and air-dried before being placed in our vacuum chamber. Mg-B precursor films 100-nm to 200-nm thick with a nominal composition of 1:2 were prepared on the unheated foil by simultaneous evaporation of Mg (purity 99.8%) and B (purity 99.9 %) from separate W and Ta resistive heaters at a vacuum of $8 \times 10^{-4}$ Pa. After the deposition, the Mg-B films were *ex-situ* annealed in a special low-temperature process to avoid overheating the Kapton foil, which had to be kept below its 300°C deformation temperature.

Several rapid annealing procedures have been implemented. Initially, the Mg-B films were placed inside a quartz tube on a thick sapphire plate and introduced for 10 to 300 s into a preheated furnace. The furnace temperature varied from 350°C to 500°C, and the Ar atmosphere could be changed from 3 kPa to 100 kPa in a flow regime. In another approach, the Mg-B films were placed "face down" on a resistive heater and annealed at 500°C to 600°C in pure Ar for 60 to 180 s. For overheating protection, the foil was covered with a sapphire plate and a large Cu block on top. After annealing, samples were cooled down within approximately 30 s to room temperature. Both of the above annealing methods resulted in superconducting $MgB_2$ with $T_{c,on}$ up to 33 K and a transition temperature $\Delta T_c \approx 10$ K. The highest $T_{c,on}$ was obtained after annealing for 180 s at 500°C on a resistive heater. Unfortunately, under such conditions, the Kapton foil changed its color from yellow to black and partially lost its flexibility. The maximal $j_c$ was only about 500 A/cm$^2$ at 4.2 K, suggesting that the damaged foil stressed the $MgB_2$ film.

Much better results were obtained using the third annealing method. Thus, the remainder of the work will be devoted to those films. The Mg-B thin films were radiatively heated with halogen lamps in vacuum; simultaneously, the substrate side of our samples was placed on an external water-cooled radiator to protect the plastic foil. The distance between the samples and the halogen source was 7 cm, and the vacuum chamber was pumped down to a base pressure of $1 \times 10^{-2}$ Pa in order to minimize the oxygen content and other gas impurities during annealing. The foil temperature was controlled by a thermocouple located very close to its surface. Time duration of the annealing process was 60 to 180 s, and temperatures at the film surface varied from 300°C to 650°C. After annealing, the vacuum chamber was filled with Ar and the films were cooled to room temperature in 20 to 30 s. Even for the highest Mg-B annealing temperature the Kapton was always maintained below 300°C, no deformation or change of its color was observed, and the resulting $MgB_2$ samples were fully flexible.

Figure 1 shows the surfaces and the critical temperature parameters of three films annealed under different conditions. For samples annealed for at least 1 min in temperatures of 550°C to 650°C at the film surface, the film morphology exhibited domains or lamellar structures [Figs. 1(a) and 1(b), respectively] and their maximal $T_{c,on}$ was only 20 K and 30 K, respectively. The film presented in Fig. 1(b) showed signs of the heat-induced substrate damage and although its $T_{c,on}$ was high, the $\Delta T_c$ was very wide. After extensive trial-and-error studies, we realized that the best annealing conditions consisted of preheating at 300°C and only very brief, 30-s heating at 600°C, followed by post-heating for 60 s again at 300°C. The films produced in this manner were characterized by very smooth surfaces [Fig. 1(c)], without domains, cracks, or lamellar structures. The x-ray diffraction measurement exhibited no diffraction peaks, thus indicating an amorphous phase in analogy with similar post-annealed $MgB_2$ films, prepared on $Si/SiO_2$ and sapphire substrates from Mg-B precursors.[6]

The homogeneity of our thin films was studied by Auger spectroscopy. The Auger spectra (Fig. 2) show a strong non-stoichiometry of Mg and B on the film surface, as well as an enhanced content of oxygen and presence of carbon. The damaged surface layer was only ~20 nm thick, however, as estimated by the rate of Mg-B etching and scanning electron microscopy studies. The bulk of the film had a uniform 1:2 stoichiometry ratio, with only small oxygen content, apparently from the residual oxygen in the annealing vacuum chamber.

The films prepared according to the annealing recipe shown in Fig. 1(c) also had the best superconducting properties. The maximal obtained $T_{c,on}$ was 29.3 K, as shown in Fig. 3, where we plotted the resistive superconducting transitions for both the stripe cut from a film by scissors and the one patterned by photolithography and Ar ion etching. The cut stripe was approximately 1-mm wide and represents the superconducting properties of our plain MgB$_2$ films, while the patterned bridge was 10-$\mu$m wide and 120-$\mu$m long. We note that Ar-ion etching resulted in a slight reduction of $T_{c,\text{on}}$; at the same time, however, $\Delta T_c$ narrowed to 2 K. The patterned microbridge was also used for $j_c$ measurements. The inset in Fig. 3 shows the $j_c$ dependence on temperature together with the fit based on the $j_c(T) = k(1-T/T_c)^\alpha$ expression, where $k$ is a constant and $\alpha$ is the fitting parameter. The fit shown in Fig. 3 was obtained for $\alpha = 2$ and indicates that our film is granular with a network of superconductor–normal metal–superconductor (SNS) weak links.[11,12] At 4.2 K, $j_c$ reached the value $>7 \times 10^5$ A/cm$^2$.

In conclusion, the preparation of the superconducting MgB$_2$ thin films on the flexible plastic foils has been presented. Our films were amorphous and exhibited a very smooth surface. The films were annealed under optimal conditions, using rapid radiative (halogen lamps) Mg-B annealing and simultaneous water-cooling of the Kapton foil, and were characterized by $T_{c,\text{on}}$ of about 29 K, $\Delta T_c$ of about 3 K, and $j_c > 7 \times 10^5$ A/cm$^2$ at 4.2 K. The measured $j_c(T)$ characteristics indicated the presence of the SNS weak-link network in our films. The Auger measurement showed that besides the ~20-nm-thick film surface, the bulk of the film exhibited a fully stoichiometric composition of Mg and B. Finally, we note that our rapid annealing procedure prevented any substrate degradation and is suitable for annealing of thin films prepared on various, unstable at high temperature substrates.

The authors thank E. Dobrocka for help in x-ray diffraction, M. Zahoran for scanning electron microscopy, and M. Gregor for the Auger spectroscopy studies. This work was supported by the Slovak Grant Agency for Science, Grants No. 2/7072/2000 and 2/7199/2000), US NSF grant DMR-0073366, and by the NAS COBASE 2000 Program, supported by the contract INT-0002341 from NSF.

**Figure Captions**

Fig. 1. Scanning-electron-microscope pictures of three $MgB_2$ thin films annealed under different conditions in the halogen lamp process. Each picture is identified by the film annealing temperatures and times, and the resulting $T_{c,on}$ and $\Delta T_c$. The optimal conditions correspond to the film shown in panel (c).

Fig. 2. The depth profile of an optimally annealed [Fig. 1(c)] $MgB_2$ film, obtained from the Auger measurement.

Fig. 3. Resistance versus temperature characteristics of an optimally annealed $MgB_2$ film (open circles) and the photolithographically patterned $MgB_2$ microbridge (closed circles). The inset shows the $j_c(T)$ dependence measured for a small microbridge, together with the theoretical fit illustrating the SNS, weak-link dependence.

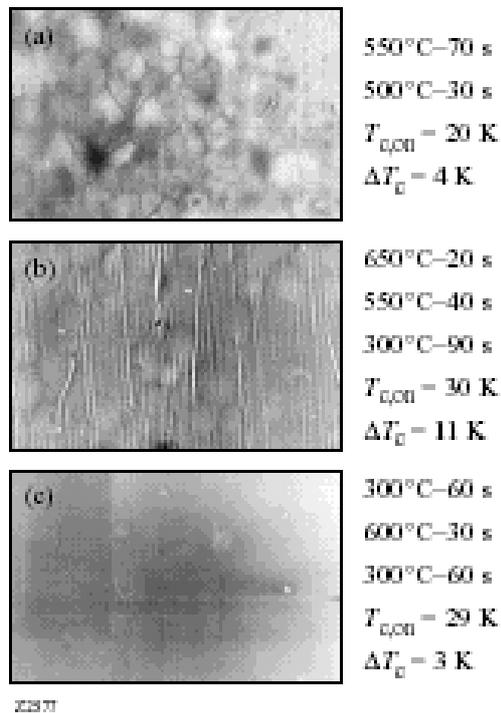

(a) 550°C–70 s
500°C–30 s
$T_{c,on} = 20$ K
$\Delta T_c = 4$ K

(b) 650°C–20 s
550°C–40 s
300°C–90 s
$T_{c,on} = 30$ K
$\Delta T_c = 11$ K

(c) 300°C–60 s
600°C–30 s
300°C–60 s
$T_{c,on} = 29$ K
$\Delta T_c = 3$ K

Fig. 1   Kus et al.

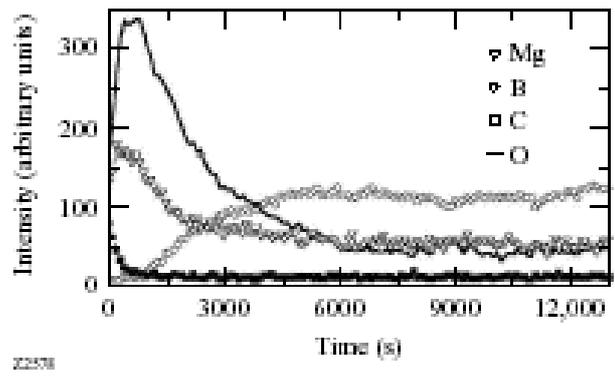

ZZ2578

Fig. 2   Kus et al.

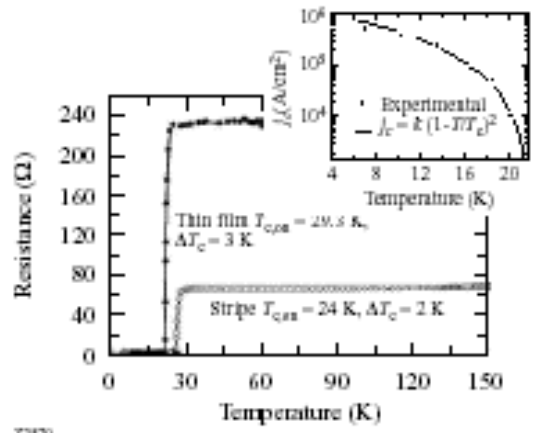

Fig. 3  Kus et al.